# Linearized Turbulent Fluid Flow as an Analog Model for Linearized General Relativity (Gravitoelectromagnetism)

H. E. Puthoff
Institute for Advanced Studies at Austin
11855 Research Blvd.
Austin, Texas 78759

## ABSTRACT

In the spirit of *analog models of and for general relativity*,[1] we explore the isomorphism between the equations of linearized turbulent fluid flow and those of the linearized form of GR, gravitoelectromagnetism. The correspondence between the truncated Reynolds system for turbulent fluid flow and the Maxwell-like gravitoelectromagnetic equations for weak-field GR is shown to provide yet another example of the relationship between the representation of fluctuations in background media and curved spacetime metric formulations.

## 1. Introduction

Recent efforts in mainstream GR studies have shown a movement toward exploring the analog relationships between GR-type phenomena (e.g., curved spacetime metrics, Riemannian geometries) and fluctuations in various background media such as fluids, condensed matter, etc. Exploration along these lines, which goes under the rubric *analog models of and for gravity*, has resulted in numerous publications - including a paper by this title [1], a book of collected essays on the topic [2], and a comprehensive web-based review with addressable links [3].

In the present literature where it has been shown that fluctuational phenomena in background media can under certain rather general conditions lead to spacetime metric representations that mimic the curved spacetime metrics of GR, the topics addressed range from models *of* gravity, e.g., the possibility of generating analog models for experiments in the laboratory concerning Hawking radiation from black holes [4], to more fundamental models *for* gravity that treat gravity as an *effective* theory arising from an underlying vacuum structure such as a quantum liquid [5].

As an additional contribution to this literature, we draw attention to publications that show that under well-defined conditions the linearized equations of *turbulent fluid flow* lead to a Maxwellian form [6]. Therefore, in contradistinction to ordinary fluid flow that supports only the propagation of longitudinal waves, turbulent fluid flow can support the propagation of transverse waves. Of interest here is that the Maxwellian form is precisely the first-order form of linearized GR, the gravitoelectromagnetic equation form by which GR takes a form isomorphic to Maxwell's equations for electromagnetism.[2] Thus it would appear that the terms appearing in basic GR equations, which are otherwise expressed in terms of metric coefficients, can now be identified with fundamental hydrodynamic variables such as averaged velocity, pressure and force density in turbulent fluid flow. Worthy of later exploration, perhaps, is whether such modeling simply provides new mathematical techniques to investigate GR-type phenomena and extensions thereof, or if in fact such isomorphism supports a conjecture that an underlying turbulent fluid structure constitutes an inherent physical aspect of nature.

---

[1] See Ref. 1.

**2. Turbulent Fluid Flow Equations**

The equations of motion for an element of a nonviscous incompressible fluid are given by Euler's hydrodynamical equations

$$\rho \frac{d\mathbf{v}}{dt} = -\nabla P \tag{1}$$

where $\rho$ is the density of the fluid, $\mathbf{v}$ its velocity, and $P$ the fluid pressure. In addition we have the equation of continuity

$$\nabla \cdot (\rho \mathbf{v}) = -\frac{\partial \rho}{\partial t} \tag{2}$$

The so-called *substantive* derivative $d\mathbf{v}/dt$ that follows an element of fluid can be expressed in terms of its spatial and temporal components as

$$\frac{d\mathbf{v}}{dt} = \frac{\partial \mathbf{v}}{\partial t} + (\mathbf{v} \cdot \nabla) \mathbf{v} \tag{3}$$

For an incompressible fluid $\rho$ is constant, and thus the above three equations can be restated in the compact form

$$\partial_t v_i + \mathbf{v} \cdot \nabla v_i + \partial_i p = 0, \qquad \nabla \cdot \mathbf{v} = 0 \tag{4}$$

where $\partial_t = \partial/\partial t$, $\partial_i = \partial/\partial x_i$, $v_i$ is the i$^{\text{th}}$ component of the velocity vector $\mathbf{v}$, and $p$ is the *specific* pressure $P/\rho$.

For a fully developed turbulent fluid it is assumed that the velocity $\mathbf{v}$ and pressure $p$ at a point take on completely random values, which can be expressed as

$$\mathbf{v} = \langle \mathbf{v} \rangle + \mathbf{v}', \qquad p = \langle p \rangle + p' \tag{5}$$

where $\mathbf{v}'$ and $p'$ are fluctuations about their respective means. Substitution into (4) followed by averaging leads to

$$\partial_t \langle v_i \rangle + \langle \mathbf{v} \rangle \cdot \nabla \langle v_i \rangle + \langle \mathbf{v}' \cdot \nabla v_i' \rangle + \partial_i \langle p \rangle = 0, \qquad \nabla \cdot \langle \mathbf{v} \rangle = 0 \tag{6}$$

where we have assumed that the averages of the fluctuations themselves vanish (i.e., $\langle \mathbf{v}' \rangle = \langle p' \rangle = 0$), and we take note of the fact that in averaging the following rules apply:

$$\left\langle \frac{\partial f}{\partial q} \right\rangle = \frac{\partial \langle f \rangle}{\partial q}, \langle \langle f \rangle \rangle = \langle f \rangle, \langle f + g \rangle = \langle f \rangle + \langle g \rangle, \langle \langle f \rangle g \rangle = \langle f \rangle \langle g \rangle, \langle f g \rangle \neq \langle f \rangle \langle g \rangle \tag{7}$$

---

[2] In the original publications the emphasis was on modeling electromagnetism in terms of the equations of turbulent fluid flow.

Following the development in Troshkin [6] we subtract (6) from (4) to obtain the intermediate equations

$$\partial_t v_i' + \langle \mathbf{v} \rangle \cdot \nabla v_i' + \mathbf{v}' \cdot \langle \nabla v_i \rangle + \mathbf{v}' \cdot \nabla v_i' - \langle \mathbf{v}' \cdot \nabla v_i' \rangle + \partial_i p' = 0 \tag{8}$$

and

$$\nabla \cdot \mathbf{v}' = 0 \tag{9}$$

Multiplication of (8) by $v_j'$ followed by exchange of $i$ and $j$ to yield a second equation, then summation of the two equations followed by averaging, yields a canonical form for turbulent fluid flow

$$\partial_t \langle v_i' v_j' \rangle + \langle \mathbf{v} \rangle \cdot \nabla \langle v_i' v_j' \rangle + \langle v_i' \mathbf{v}' \rangle \cdot \langle \nabla v_j \rangle + \langle v_j' \mathbf{v}' \rangle \cdot \langle \nabla v_i \rangle + h_{ij} = 0 \tag{10}$$

where $h_{ij} = \langle v_i' \partial_j p' \rangle + \langle v_j' \partial_i p' \rangle + \nabla \cdot \langle v_i' v_j' \mathbf{v}' \rangle$. \qquad (11)

The significance of various terms in the above two equations are as follows. The average of the product of velocity fluctuations, $\langle v_i' v_j' \rangle \triangleq \tau_{ij}$ form a symmetric matrix called the Reynolds tensor whose nonzero value is indicative of a turbulent medium. (The Reynolds tensor = 0 for nonfluctuating laminar media.) The terms in $h_{ij}$ correspond to the elements of turbulent relaxation (first two terms) and turbulent diffusion (last term). Given the complexity of generating analytical solutions for these terms they are typically treated on the basis of phenomenological equations which we shall do here. For example, following Troshkin we shall neglect the diffusion of turbulence in comparison with its relaxation, and thus assume that the last term $\nabla \cdot \langle v_i' v_j' \mathbf{v}' \rangle = 0$ [6]. For turbulent relaxation we assume the sum of the first pair of terms can be represented by an approximate form

$$h_{ij} = \alpha \left( \tau_{ij} - \tau_{ij}^{(0)} \right) \tag{12}$$

where the $\tau_{ij}^{(0)}$ is taken to constitute a persistent background distribution of Reynolds tensor elements of an ideal turbulent fluid, and $\alpha$ is a constant, $\alpha > 0$.

Therefore, we take as our foundational equations for a linearized turbulent fluid the equations (6), (10), and (12) written in the form of what is called a *truncated Reynolds system*

$$\partial_t \langle v_i \rangle + \langle \mathbf{v} \rangle \cdot \langle \nabla v_i \rangle + \nabla \cdot \boldsymbol{\tau}_i + \partial_i \langle p \rangle = 0 \tag{13}$$

$$\partial_t \tau_{ij} + \langle \mathbf{v} \rangle \cdot \nabla \tau_{ij} + \boldsymbol{\tau}_i \cdot \langle \nabla v_j \rangle + \boldsymbol{\tau}_j \cdot \langle \nabla v_i \rangle + h_{ij} = 0 \tag{14}$$

$$\nabla \cdot \langle \mathbf{v} \rangle = 0 \tag{15}$$

$$h_{ij} = \alpha \left( \tau_{ij} - \tau_{ij}^{(0)} \right) \tag{16}$$

In the above we have introduced the notation $\tau_{ij} \triangleq \langle v_i' v_j' \rangle$, $\boldsymbol{\tau}_i \triangleq \langle v_i' \mathbf{v}' \rangle$, and noted that (following from (9)) $\langle \mathbf{v}' \cdot \nabla v_i' \rangle = \nabla \cdot \langle v_i' \mathbf{v}' \rangle = \nabla \cdot \boldsymbol{\tau}_i$.

## 3. Application to Development of GR Gravitoelectromagnetic Equations

We now explore the isomorphism between the equations of linearized turbulent fluid flow and those of gravitoelectromagnetism, the linearized form of GR. We begin with an ideal turbulent fluid in a dynamical state $\langle \mathbf{v} \rangle, \langle p \rangle$ and $\tau_{ij}$ which is close to persistent background values $\langle \mathbf{v}^{(0)} \rangle, \langle p^{(0)} \rangle$ and $\tau_{ij}^{(0)}$ which satisfy equations (13) – (16). For small disturbances from the background we substitute into (13) - (16)

$$\langle v_i \rangle = \langle v_i^{(0)} \rangle + \langle \Delta v_i \rangle, \quad \langle p \rangle = \langle p^{(0)} \rangle + \langle \Delta p \rangle, \quad \tau_{ij} = \tau_{ij}^{(0)} + \Delta \tau_{ij} \tag{17}$$

In the process we neglect product terms quadratic in the disturbances, assume a stationary average background fluid velocity such that $\langle \mathbf{v}^{(0)} \rangle = 0$, and assume for a homogeneous, isotropic background $\tau_{ij}^{(0)} = c^2 \delta_{ij}$, where $\delta_{ij} = 1$ for $i = j$, 0 otherwise, and c is a constant. The result is

$$\partial_t \langle \Delta v_i \rangle + \nabla \bullet (\Delta \tau_i) + \partial_i \langle \Delta p \rangle = 0 \tag{18}$$

$$\partial_t \Delta \tau_{ij} + c^2 \left( \partial_i \langle \Delta v_j \rangle + \partial_j \langle \Delta v_i \rangle \right) + h_{ij} = 0 \tag{19}$$

$$\nabla \bullet \langle \Delta \mathbf{v} \rangle = 0 \tag{20}$$

$$h_{ij} = \alpha \, \Delta \tau_{ij} \tag{21}$$

We now introduce (i.e., define) gravitoelectromagnetic (G.E.M.) variables of interest by assigning the following relationships, where $K$ is a proportionality constant:

$$\left( E_g \right)_i = K \nabla \bullet \Delta \tau_i \tag{22}$$

$$\mathbf{A}_g = 2K \langle \Delta \mathbf{v} \rangle \tag{23}$$

$$\mathbf{B}_g = 2K \nabla \times \langle \Delta \mathbf{v} \rangle \tag{24}$$

$$\varphi_g = K \langle \Delta p \rangle \tag{25}$$

$$\rho_M = -\frac{K}{4\pi G} \partial_i \partial_j \Delta \tau_{ij} \tag{26}$$

$$\left( J_M \right)_j = -\frac{K}{4\pi G} \nabla \bullet \mathbf{h}_j \tag{27}$$

where, from (21), $\mathbf{h}_j = \alpha \Delta \mathbf{\tau}_j$.

From (23) and (24) we obtain the first of the G.E.M. equations of interest [7,8]

$$\left( \mathbf{B}_g / 2 \right) = \nabla \times \left( \mathbf{A}_g / 2 \right) \tag{28}$$

By vector identity the divergence of (28) yields a second G.E.M. equation

$$\nabla \cdot (\mathbf{B}_g / 2) = 0 \tag{29}$$

Multiplication of (18) by $K$ followed by substitution of G.E.M. variables defined above yields a third G.E.M. equation

$$\mathbf{E}_g = -\nabla \varphi_g - \frac{\partial (\mathbf{A}_g / 2)}{\partial t} \tag{30}$$

Curl of the above then leads to a fourth G.E.M. equation

$$\nabla \times \mathbf{E}_g = -\frac{\partial (\mathbf{B}_g / 2)}{\partial t} \tag{31}$$

Application of $\partial_i$ to (22) followed by use of the definition of $\rho$ given in (26) yields a fifth G.E.M. equation

$$\nabla \cdot \mathbf{E}_g = -4\pi G \rho_M \tag{32}$$

To obtain the sixth G.E.M. equation the following steps are required: Eq. (19) is multiplied by $K$, then followed by the operation of $\partial_i$ on (19) to obtain the vector equation

$$\partial_t \nabla \cdot K \Delta \boldsymbol{\tau}_j + c^2 \nabla^2 \left( K \langle \Delta \mathbf{v} \rangle \right)_j + c^2 \partial_j K \nabla \cdot \langle \Delta \mathbf{v} \rangle + K \nabla \cdot \mathbf{h}_j = 0 \tag{33}$$

Application of the vector identity $\nabla^2 \mathbf{a} = \nabla \nabla \cdot \mathbf{a} - \nabla \times \nabla \times \mathbf{a}$ and use of Eq. (20) followed by substitution of the G.E.M. variables then yields the sixth G.E.M. equation

$$\nabla \times (\mathbf{B}_g / 2) = -\frac{4\pi G}{c^2} \mathbf{J}_M + \frac{1}{c^2} \frac{\partial \mathbf{E}_g}{\partial t} \tag{34}$$

We also note that (20) and (23) together define the Coulomb gauge

$$\nabla \cdot (\mathbf{A}_g / 2) = 0 \tag{35}$$

a consequence of the assumed incompressibility of the ideal turbulent fluid.

If the incompressibility assumption is relaxed to accommodate a slight compressibility, i.e., $\rho = \langle \rho \rangle + \rho'$, where $\rho'$ is a fluctuational component, the Coulomb gauge can be shown to transform into the Lorentz gauge form as follows [9]. In this case, (2), rather than reducing to (15), takes the form (following the averaging procedure)

$$\nabla \cdot (\langle \rho \rangle \langle \mathbf{v} \rangle) = -\partial_t \langle \rho \rangle \tag{36}$$

where, as in the case of the velocity and pressure terms, the average fluctuational component of the density $\langle \rho' \rangle = 0$, and we take the average of a product term quadratic in perturbations, $\langle \rho' \mathbf{v}' \rangle$, to be negligible. Then, as in the case of the velocity and pressure terms, we assume that the average density remains close to a persistent background value $\langle \rho^{(0)} \rangle$ such that

$$\langle \rho \rangle = \langle \rho^{(0)} \rangle + \langle \Delta \rho \rangle \tag{37}$$

For small disturbances (again, neglecting quadratic terms) (36) reduces to

$$\langle \rho^{(0)} \rangle \nabla \cdot \langle \Delta \mathbf{v} \rangle = -\partial_t \langle \Delta \rho \rangle \tag{38}$$

Now we note that, in general for a fluid, the velocity-squared of the propagation of a disturbance (in this case a transverse wave of turbulence perturbation propagating as a compound density-turbulence perturbation wave), which we take to be the velocity of light, is given in terms of a ratio of the average pressure perturbation to average density perturbation. Thus

$$v^2 = c^2 = \frac{\langle \Delta P \rangle}{\langle \Delta \rho \rangle} = \frac{\langle \Delta (\langle \rho^{(0)} \rangle p) \rangle}{\langle \Delta \rho \rangle} \tag{39}$$

Substitution of (39) into (38) followed by insertion of the G.E.M.-defined variables then yields the Lorentz gauge condition

$$\nabla \cdot (\mathbf{A}_g / 2) = -\frac{1}{c^2} \frac{\partial \varphi_g}{\partial t} \tag{40}$$

Based on this derivation we find that the Coulomb and Lorentz gauges correspond, respectively, to the fluid flow conditions of incompressible vs. slightly compressible turbulence.

Therefore, the first-order linearized form of GR, the gravitoelectromagnetic equations, can be seen as isomorphic to the equations of a first-order linearized form of turbulent fluid flow, the so-called "truncated" Reynolds system. Of additional note is the fact that although the fundamental equations of fluid flow were nonrelativistic in nature, the resulting Maxwell-form equations for turbulent fluid flow are Lorentz-invariant.

Therefore in this ansatz Lorentz-invariance is an *emergent* property associated with the dynamics of an underlying turbulent fluid flow.

## 4. Spacetime Metric as a Function of Turbulent Fluid Parameters

With the above derivations in place, one can now express the metric coefficients $g_{\mu\nu}$ in terms of the hydrodynamic variables of the turbulent fluid flow model. In the G.E.M. formalism the metric coefficients are given by $g_{\mu\nu} = \eta_{\mu\nu} + h_{\mu\nu}$, where $\eta_{\mu\nu}$ is the diagonal Minkowski metric with signature (+1, -1, -1, -1). In terms of the definition of the G.E.M. variables, $(\varphi_g, \mathbf{A}_g)$, $g_{\mu\nu}$ is given by [8],

$$g_{00} = 1 + \frac{2\varphi_g}{c^2}; \quad g_{0i} = \frac{2A_i}{c}; \quad g_{ij}(i \neq j) = 0; \quad g_{11} = g_{22} = g_{33} = -1 + \frac{2\varphi_g}{c^2} \quad (41)$$

Substitution of definitions (23) and (25) then yields

$$g_{\mu\nu} = \begin{pmatrix} 1 + 2K\langle\Delta p\rangle/c^2 & 4K\langle\Delta v_x\rangle/c & 4K\langle\Delta v_y\rangle/c & 4K\langle\Delta v_z\rangle/c \\ 4K\langle\Delta v_x\rangle/c & -1 + 2K\langle\Delta p\rangle/c^2 & 0 & 0 \\ 4K\langle\Delta v_y\rangle/c & 0 & -1 + 2K\langle\Delta p\rangle/c^2 & 0 \\ 4K\langle\Delta v_z\rangle/c & 0 & 0 & -1 + 2K\langle\Delta p\rangle/c^2 \end{pmatrix} \quad (42)$$

## 5. Conclusions

Based on the identifications assigned by (22) – (27) one infers that the gravitational scalar potential $\varphi_g$, vector potential $\mathbf{A}_g$, gravitomagnetic field $\mathbf{B}_g$, and gravitoelectric field $\mathbf{E}_g$ correspond to averages of departures from an ideal, isotropic, homogeneous, nearly incompressible background turbulent fluid of, respectively, pressure, velocity, vorticity and force density. What remains to be determined by further study are three issues: (1) whether the isomorphism between the equations of linearized turbulent fluid flow and those of linearized GR simply provides for a mechanical modeling of gravitoelectromagnetism or has a deeper significance with regard to an underlying reality of some form of vacuum fluid structure [5], (2) whether appropriate extension of the linearized turbulent fluid flow analysis into the nonlinear regime can reproduce the nonlinear aspects of GR in detail, and (3) whether yet further extension of the mathematical development of fluid flow turbulence can uncover additional features that encompass spacetime structures characteristic of other fields and their associated particles [10].

**Acknowledgements**  I would like to thank M. Derakhshani for bringing key references to my attention, and to express my appreciation to M. Ibison and G. Hathaway for valuable inputs during the preparation of this paper. I'm also indebted to F. Winterberg for his original work and helpful correspondence concerning the themes considered in this effort.